\documentclass[12pt]{article}

\usepackage{latexsym}

\font\tenmsbm=msbm10 scaled 1200
\font\sevenmsbm=msbm9
\newfam\msbmfam
\textfont\msbmfam=\tenmsbm
\scriptfont\msbmfam=\sevenmsbm
\def\msbm{\fam\msbmfam\tenmsbm}

\makeatletter
\@addtoreset{equation}{section}
\makeatother
\renewcommand{\theequation}{\thesection.\arabic{equation}}
\newcommand{\eref}[1]{(\ref{#1})}

\def\beq{\begin{equation}}
\def\eeq{\end{equation}}
\def\bea{\begin{eqnarray}}
\def\eea{\end{eqnarray}}
\def\bet{\begin{tabular}}
\def\eet{\end{tabular}}
\def\ol{\overline}
\def\nin{\noindent}
\def\a{\alpha}
\def\b{\beta}
\def\l{\lambda}
\def\f{\phi}
\def\vf{\varphi}
\def\c{\chi}
\def\g{\gamma}
\def\s{\sigma}
\def\e{\epsilon}
\def\G{\Gamma}
\def\tF{\tilde{F}}
\def\tH{\tilde{H}}
\def\tV{\tilde{V}}
\def\tB{\tilde{B}}
\def\tA{\tilde{A}}
\def\tf{\tilde{\f}}

\def\D{{\cal{D}}}
\def\R{{\cal{R}}}

\begin{document}

\begin{titlepage}

\begin{flushright}
Preprint DFPD 97/TH/27\\
hep-th/97xxxxx\\
July 1997\\
\end{flushright}

\vspace{2truecm}

\begin{center}

{\bf \Large $N = 1, D = 6$ Supergravity: Duality and non Minimal 
Couplings.}

\vspace{1cm}

{ Gianguido Dall'Agata} and { Kurt 
Lechner}\footnote{kurt.lechner@pd.infn.it}

\vspace{1cm}

{\it Dipartimento di Fisica, Universit\`a degli Studi di Padova,

\smallskip

and

\smallskip

Istituto Nazionale di Fisica Nucleare, Sezione di Padova, 

Via F. Marzolo, 8, 35131 Padova, Italia}

\vspace{1cm}

\begin{abstract}

Six-dimensional supergravity theories and their duality properties play a
central role in the context of string duality and string compactifications.
Lowering dimensions leads usually to an increasing complexity of theories;
with this respect six dimensions seem to constitute an appropriate
compromise between the physical four and the presumably more fundamental
ten or eleven dimensions. In this paper we present a superspace formulation
of $N = 1, D = 6$ supergravity with one tensor-multiplet and an arbitrary
number of vector- and hypermultiplets, in which the
bosonic abelian superforms of the theory, the dilaton, the abelian gauge
fields and the two-form are replaced by their S-duals i.e. four, three and
two-superforms respectively, in compatibility with supersymmetry. As usual
this replacement interchanges Bianchi identities with equations of motion.
This formulation holds in the presence of one tensor multiplet and
arbitrary numbers of hypermultiplets and abelian  super-Maxwell multiplets
if all couplings are minimal. We determine the  consistency conditions for
non-minimal couplings in $N = 1$, $D = 6$  supergravity, for which we
present a particularly significant solution, namely the one associated with
the Chern-Simons-Lorentz three-form which entails the Green-Schwarz
anomaly cancellation mechanism. In the case of non minimal couplings it is
found that the gauge fields and the two-form can still be dualized while
the dilaton has to remain a zero-form.

\end{abstract}

\end{center}
\end{titlepage}

\newpage

\baselineskip 6 mm

%%%%%%%%%%%%%%%%%%%%%%%%%%%%%%%%%%%%%%%%%%%%%%%%%%%%%%%%%%%%%%%%%%%
%%%%%%%%%%%%%%%%%%%%%%%%%%%%%%%%%%%%%%%%%%%%%%%%%%%%%%%%%%%%%%%%%%%

\section{Introduction}

Borne as an attempt of quantizing gravity, supergravity theories in 
diverse dimensions play nowadays an important role as low-energy 
effective  field theories of superstring and membrane theories.  
Supergravity theories have been extensively studied in four 
dimensions, of course because of their direct physical relevance, and 
in ten and eleven dimensions because of their fundamental features.

Even if it is known that consistent  superstring theories can 
be formulated in six dimensions and that six-dimensional supergravity 
can arise as their low-energy limit, this is not the only reason for
investigating  $D=6$ supergravity.
In fact, while $D = 10$ and $D = 11$ theories are of direct interest as 
backgrounds for strings, membranes and M-theory, one frequently performs 
compactifications down to $D = 6$ to clarify relations among these
theories, which are hidden in their ten or eleven-dimensional formulations.

Actually, the greatest recent step forward in understanding superstring 
theories was the discovering of {\it duality}. 
All known superstring theories, membrane theories and 
M-theory are related by duality and can be regarded as
expansions of a unique theory around different vacua. 
Most of the times these relations are evident only after compactification,
frequently to six dimensions.

For example, it has been conjectured \cite{0} that compactification of 
the type IIA theory on $K_3$  and the heterotic string on ${\msbm T}^4$ 
gives rise to the same theory on the remaining six-dimensional space
and in \cite{SM} it has been shown  heuristically that this equivalence can be 
understood as a consequence of M-theory.
Another example is M-theory 
compactified on $ (K_3 \times S_1)/ {\msbm Z}_2$ (see \cite{Sen}) which
gives rise to new couplings in $N = 1,D=6$  supergravity which have been
studied in \cite{NS}.

Moreover, recently a possible new fundamental 
theory (F-theory) living in a twelve--dimensional space with signature $(10,2)$
has been conjectured in \cite{Vafa}.
As shown in \cite{Vafa}, compactifications of F-theory on Calabi-Yau manifolds
should give rise to supergravity theories in $D=6$ with $N=1$ supersymmetry. 
Some details of these compactifications are available in \cite{MF}.

For more details on the relevance  of  six--dimensional supergravity 
theories see for example the introduction in \cite{BSS}.

In this paper we are concerned with the Hodge S-duality of $p$--forms in 
six-dimensional supergravity addressing the problem of the dualizability
of these forms in compatibility with supersymmetry.
While for $D = 11$ supergravity this analysis has been performed in 
\cite{CL, Torinesi}, for ten dimensions in \cite{LT} the authors 
presented a formulation of $N=1, D=10$ 
supergravity super-Maxwell theory in superspace in which {\it all the 
bosonic abelian fields} can be described as $p$-superforms or as 
$(D-p-2)$-superforms, the bosonic components of the supercurvatures of these 
fields being related by Hodge-duality. For the corresponding 
results in $IIA$ and $IIB$ supergravity see ref. \cite{Ceder}.
The conjecture that emerges from these papers is that in every 
Supergravity theory and in every dimension an abelian form can 
be described alternatively in either of the 
two ways in compatibility with supersymmetry.

The results of the present paper extend the validity of
this conjecture also to  the six--dimensional case. 
Actually, these features hold true if the couplings are all
minimal. Interesting non minimal couplings arise (in two, six and
ten dimensions) from a Lorentz--Chern--Simons term in the two-form
curvature, which realizes the Green--Schwarz anomaly cancellation
mechanism.
A supersymmetric version of this mechanism \cite{BPT,catena,Rid} amounts to a 
supersymmetrization of the Lorentz-Chern-Simons form
or equivalently to the solution of the superspace Bianchi identity
\beq
d H = c_1 tr F^2 + c_2 tr R^2 \label{CSL}
\eeq
for $c_2 \neq 0$. This problem has been solved in ten dimensions 
through the so called Bonora-Pasti-Tonin theorem \cite{BPT}.
In this paper we prove the six--dimensional version of this theorem
which guarantees essentially the solvability of \eref{CSL}
in superspace. 
This allows us to test the duality conjecture also in the presence of 
the non-minimal couplings arising from the $tr R^2$ term in
\eref{CSL}. The principal result we found is that all the abelian forms 
but the dilaton can still be dualized. 

In this paper we deal with $N=1, D=6$ supergravity with 
one tensor multiplet, an arbitrary number of vector multiplets and 
hypermultiplets in a superspace formalism \cite{N,old,NG,susy6d}.
In section two we present our superspace conventions and notations. 

Section three is devoted to the solution of the relevant superspace
Bianchi identities (B.I.). First of all
we perform a group theoretical analysis of the 
superspace constraints. This  allows us first to classify them and, 
second, to derive consistency conditions on the  remaining 
undetermined auxiliary fields which trigger the couplings between the 
multiplets: each choice for the auxiliary fields satisfying these 
consistency conditions gives rise to a different theory.

A more precise analysis of these consistency conditions shows that 
even with minimal couplings we are 
not dealing with a unique theory but with a {\it family of theories}. 
For these theories the couplings depend on a real parameter $k$ which 
cannot be scaled away. As one sends $k \to 0$ one obtains the 
standard minimal couplings between all the multiplets, while for $k 
\to \infty $ one obtains the coupled tensor + Yang-Mills system in flat 
space, i.e. the supergravity multiplet decouples. This should present 
the limiting procedure missed in \cite{BSS}.

Having solved the fundamental B.I.'s, in section four we 
construct superforms dual respectively to the dilaton, Maxwell gauge 
fields, and the two-form $B$, such that they satisfy dual B.I.'s in
superspace. These 
represent the equations of motion for the basic abelian forms and allow, in 
turn, to define dual super-potentials, i.e. four, three and 
two-superforms respectively. The basic B.I.'s can then be 
interpreted as equations of motion for the dual potentials.

In this way we give a strong support to the 
conjecture that every abelian $p$-superform in a supergravity 
theory can indeed be described by a dual $(D-p-2)$-superform 
although a general proof of this conjecture is still missing.

In section four the analysis was performed for simplicity in the absence 
of hypermultiplets, so in section five we extend the above mentioned 
results to the case where the hypermultiplets are present. It remains 
an open question whether the scalars belonging to the hypermatter can 
themselves be transformed to four-forms (see however \cite{susy6d}).

Section six is devoted to the supersymmetrization of the 
Lorentz--Chern--Simons form, which presents a non--minimal solution
of the above mentioned consistency conditions, and to an analysis
of the duality properties of the resulting theory.

Section seven contains some concluding remarks while in the appendix
we give more details on our notations and a few fundamental gamma
matrix identities.

\section{Preliminaries}

Six-dimensional $N = 1$ supergravity allows for four kind of 
multiplets:
\bea
\hbox{Pure supergravity} & & \{e_m^a, \psi_m^{\a i}, B^-_{ab}\} , \\
\hbox{Tensor multiplet} & & \{ B^+_{ab}, \l_{\a i}, \f \}, \\
\hbox{Yang-Mills} & & \{ A_a, \c^{\a i}\}, \\
\hbox{Hypermultiplet} & & \{ \psi^{Y}_{\a} , \vf^{I}\} .
\eea

The superspace in six dimensions is spanned by the supercoordinates 
$Z^M =  (x^m, \vartheta^{\mu i})$ where $x^m$ ($m = 
0,\ldots,5$) are the 
ordinary space-time coordinates and $\vartheta^{\mu i}$ ($\mu = 1, \ldots 
4 $) are symplectic Majorana-Weyl spinors carrying the $USp(1)$ doublet 
index $i = 1,2$. In what follows letters from the middle of 
the alphabet represent curved indices while letters from the beginning 
represent flat indices: small Latin letters $a = (0,\ldots, 
5)$ indicate vectorial indices, small Greek letters $\a = 
(1,\ldots,4)$ indicate spinorial indices and 
capital letters denote both of them $A = (a, \a i)$.
The fields $\phi^I$ ($I = 1, \ldots, 4n_H$) constitute 
the coordinates of the 
quaternionic K\"ahler manifold $ USp(n_H, 1) / USp(n_H) \otimes 
Usp(1)$, where $n_H$ is the number of hypermultiplets,
and the index $Y = 1,\ldots, 2n_H$ stands  for the 
fundamental representation of $USp(n_H)$.

The superspace geometry is described by the vielbeins $E^A  
= dZ^M \ {E_{M}}^A(Z)$, 
the Lorentz--valued connection $\Omega_A{}^B$, the Lie-algebra-valued 
Yang-Mills (YM) connection $A$, which are one-superforms, and the 
two-superform $B$. 
We remember that a superfield ${\psi_A}^B$ is Lorentz-valued if 
${\psi_a}^{b} = - \psi^{b}{}_a$, $\displaystyle {\psi_\a}^\b = 
\frac{1}{4} \, (\G^{ab})_\a{}^\b \; \psi_{ab}$ and vanishes otherwise.
Since our spinors are four component Weyl spinors we use
a Weyl algebra of  $4 \times 4$ $\Gamma$ matrices
\beq
(\G^a)_{\a\b} (\G^b)^{\b\g} + (\G^b)_{\a\b} (\G^a)^{\b\g}  = 2 \;
\eta^{ab} \, \delta^\g_\a.
\eeq
$\G^{a_1\ldots a_n}$ will denote completely antisymmetrized product 
of $\G$ matrices, where antisymmetrization is understood with 
unit weight.

A $p$-superform can be decomposed in the vielbein 
basis as
\beq
\f_p = \frac{1}{p!} E^{A_1} - E^{A_p} \f_{A_p - A_1}(Z).
\eeq

\nin
It will be useful to call $(q, p-q)$ sector of $\f_p$, denoted by 
$\f_{(q,p-q)}$, the component of $\f_p$ proportional to $q$ vector-like 
supervielbeins $E^a$ and $p-q$ spinor-like supervielbeins $E^{\a i}$.

The torsion $T^A$ and the Lorentz and Yang-Mills 
curvatures ${R_A}^B$ and $F$ are defined as 
\beq
T^A = D E^A = dE^A+ E^B\Omega_B{}^A =\frac{1}{2} E^B E^C {T_{CB}}^A,
\eeq

\beq
{R_A}^B  = d {\Omega_A}^B + {\Omega_A}^C {\Omega_C}^B  = \frac{1}{2} 
E^C E^D {R_{DCA}}^B ,
\eeq

\beq
F = dA + AA = \frac{1}{2} E^{B} E^C F_{CB},
\eeq

\nin
while the $B$-field strength $H$  depends on the model.
In general one sets
\beq
H = d B + c_1 \omega_{3YM} + c_2 \omega_{3L},
\eeq

\nin
where $\omega_{3YM}$ is the YM Chern-Simons 3-super-form which couples 
Yang-Mills to the supergravity and tensor multiplet, while  $\omega_{3L}$ 
is the Lorentz Chern-Simons form:
\beq
d \omega_{3YM} = tr F^2, \qquad d \omega_{3L} = tr 
R^2.
\eeq

\nin
We define the Hodge duality relation between 
tensors as
\beq
\tilde{W}_{a_1\ldots a_k} \equiv 
\frac{(-1)^{\frac{k}{2}(k+1)}}{(6-k)!}\e_{a_1\ldots a_6} W^{a_{k+1} 
\ldots a_6}
\eeq

\nin
which  allows to decompose antisymmetric 
three-tensors in self-dual ($+$) and anti-selfdual ($-$) parts
\beq
\label{Hodge}
W^{(\pm)}_{abc} \equiv \frac{1}{2} ( W_{abc} \pm \tilde{W}_{abc}),
\eeq

\nin
which fulfill the $\G$-projections:
\beq
(\G^{abc})_{\a\b} \; W^-_{abc} = 0, \qquad
(\G_{abc})^{\a\b} \; W^{+ abc} = 0.
\eeq

Other notations we use in what follows are $[\ldots)$, $[\ldots]$,$(\ldots)$ 
on indices denoting graded symmetrization, antisymmetrization and symmetrization 
respectively, with unit weight.

\section{The Bianchi Identities}

The super--torsion  and super--curvatures satisfy the Bianchi identities 
\bea
D T^A &=& E^B {R_B}^A, \label{IbT}\\
D {R_A}^B &=& 0, \\
D F &=& 0,\label{IbF}\\
d H &=& c_1 tr(F^2) + c_2 tr(R^2). \label{IbH}
\eea
In this section we set $c_2 = 0$ and consider $c_2 \neq 0$ in 
section six where we discuss the supersymmetrization of the 
Lorentz--Chern--Simons form.

The unphysical fields present in the supercurvatures are eliminated
by imposing suitable constraints. Once these constraints are imposed
the B.I.'s are no longer identities and have to be solved consistently.
One consistency requirement is the closure of the supersymmetry (SUSY)
algebra:
\beq
D_A D_B - (-1)^{AB} D_B D_A = - {T_{AB}}^C D_C - R_{AB\#}{}^\# - 
[F_{AB}, ). \label{algSUSY}
\eeq

\subsection{Constraint analysis}

In the superspace language the dynamics of the physical fields is 
governed by the constraints one imposes on the supercurvatures.
To classify the possible consistent sets of constraints for the 
torsion we adopt a general strategy which is based on group 
theoretical arguments, see e.g. \cite{CL,Rid}.

Our starting point is the fundamental rigid supersymmetry preserving 
constraint
\beq
T_{\a i \b j}{}^a = -2 \e_{ij} \G^a_{\a\b} .
\eeq
Once this constraint has been fixed the remaining ones fall into two 
classes: constraints which can be obtained by superfield redefinitions 
(kinematical constraints) and constraints which determine the dynamics 
and the couplings between the fields (dynamical constraints).
The basic redefinitions are \cite{Rid}
\beq
\begin{array}{c}
E^{\prime \a i} = E^\a + E^b H_b{}^{\a i}, \\
\Omega_{\a i a}^\prime{}^b = \Omega_{\a i a}{}^b + X_{\a i a}{}^b,
\end{array}
\eeq
where $ H_b{}^{\a i}$ and $X_{\a i a}{}^b$ are suitable covariant 
superfields with $X_{\a i ab}=-X_{\a i ba}$. 

With these redefinitions we can eliminate from $T_{\a i a}{}^b$ all  
irreducible representations (irrep) of $SO(1,5)$ apart from the 
$\ol{60}$ which is symmetric in its vectorial indices $ab$ and has all 
its $\G$-traces vanishing. At this point the lowest sector of the torsion 
B.I.'s 
\beq
{T_{\a i \b j}}^{\s l} {T_{\s l \g k}}^a + {T_{\a i \b j}}^b {T_{b \g 
k}}^a + (\hbox{cyclic permutations}) = 0,
\eeq

\nin
implies that also $\ol{60}$ vanishes  and that ${T_{\a i \b j}}^{\g k}$ can 
at most contain three spinors, so
\bea
\label{vincbase1}
{T_{\a i a}}^b &=& 0, \\
\label{vincbase2}
{T_{\a i \b j}}^{\g k} &=& \delta_\a{}^\g {\omega}_{\b 
i,j}{}^k + (\a i \leftrightarrow \b j),
\eea

\nin
where ${\omega}_{\b i,jk}$, symmetric in $j$ and $k$, describes the 
three spinorial degrees of freedom left. 

The choice for $\omega$ corresponds to a dynamical constraint 
\cite{NG}: setting it to zero amounts to a decoupling of the 
hypermultiplets, while identifying $\omega$ with a $USp(1)$ 
connection and promoting the global $USp(1)$ invariance of the SUSY 
algebra in $D = 6$ to a local one leads to a natural description
\cite{N} of 
the self-interactions between the hypermultiplets on a quaternionic 
manifold. Since we discuss hypermultiplets in section five, here we 
set $\omega_{\a i, jk} = 0$.

The next step is to solve the B.I.'s with canonical dimension $d = 
1$. These B.I.'s admit a unique solution modulo the connection shift
\beq
\label{ridvec}
\Omega^{\prime}_{ab}{}^c = \Omega_{ab}{}^c + Y_{a,b}{}^c
\eeq

\nin
where $Y_{a,bc} = Y_{a[bc]}$ which allows to set the torsion 
$T_{ab}{}^c$ equal to an arbitrary tensor.

For a suitable ${Y_{a,b}}^c$ one obtains as general solution for the 
dimension 1 B.I.'s:
\bea
T_{abc}  &=& T^-_{abc}, \label{ch1}\\
{T_{a \a i}}^{\b}_j &=& 3 {\delta_{\a}}^\b V_{a ij} + {\G_{ac \a}}^\b 
V^c_{ij} - \e_{ij} \frac{1}{4} {\G^{bc}}_\a{}^\b T^-_{abc}, \\
{R_{\a i\b j ab}} &=& -4 \G_{abc\,\a\b} V^c_{ij}, \label{ch3}
\eea

\nin
where $T_{abc}^-$ is a completely antisymmetric anti-selfdual tensor 
and the vector $V_{a\, ij}$ is symmetric in $i$ and $j$ and 
constitutes an auxiliary field, which will play a central role in 
what follows.

The peculiarity of the choice \eref{ch1}-\eref{ch3} lies in the fact 
that for {\it pure} supergravity one has $V_{a\, ij} = 0$ (dynamical 
constraint), i.e.
\beq
{R_{\a i \b j a}}^b = 0, \label{relq}
\eeq

\nin
in analogy with the standard Yang-Mills constraint $F_{\a i \b j} = 
0$. The relation \eref{relq} allows, moreover, to apply a general 
cohomological technique developed in \cite{Anom} for the 
determination of the (curved) supersymmetry anomalies which are 
inevitably associated to the ABBJ-$SO(1,5)$ local Lorentz anomalies 
in supersymmetric six-dimensional theories with chiral fermions (or 
bosons).

The $d = 3/2$ B.I.'s imply (and are solved by) the following three 
relations:
\bea
{R_{a\a i b c}} &=&  - 2 \G_{a\,\a\b} {T^\b_{bc i}} + 
{\G_{[ab\,\a}}^\b \left( \frac{2}{3} \e^{hk} {\G_{c]s \b}}^\rho D_{\rho h} 
 V_{ki}^s - \frac{4}{3} \e^{hk} D_{\b h} V_{c] ki}\right),\\
\G^s_{\a\b} {T_{sa}}{}^\b_i &=& \e^{hk} D_{\a h} V_{a \, ki} - 
\frac{1}{3} \e^{hk} \G_{as\,\a}{}^\b D_{\b h} V^s_{ki}Ê\\
D_{\a (i} V^a_{jk)} &=& 0. \label{18}
\eea

The first relation parametrizes $R_{a \a i, bc}$ in terms of the 
gravitino field-strength $T_{ab}{}^{\a i}$ and of the spinorial 
derivatives of $V_{a\,ij}$ and the second one is the equation of
motion for the gravitino. The third relation constitutes a 
consistency condition on the auxiliary field $V_{a\,ij}$: every 
choice one makes for $V_{a\,ij}$, implementing thus a dynamical 
constraint, has to satisfy \eref{18} identically.
A part from the trivial choice $V_{a\, ij} = 0$, which leads to pure 
supergravity (SUGRA), we will present solutions for \eref{18} 
in the present and the next 
sections.

The solution for the $F$-B.I. proceeds as follows.
The constraints on the lowest components of the Yang-Mills supercurvature 
$F$ are the usual 
constraints proposed by Nilsson for the free theory \cite{Nilsson}
\beq
F_{\a i\b j} = 0.
\eeq

\nin
Solving the $F$-B.I. with these constraints one finds that
\beq
F_{a \a i} = - 2 \;\G_{a\,\a\b} \; \c^\b_i,
\eeq

\nin
where $\c^{\a i}$ is the gluino superfield, together with the 
supersymmetry transformation rules for the gluon 
field-strength $F_{ab}$ and the gluino
\bea 
D_{\a i} F_{ab} &=& 4 \G_{[b \ \a\rho} D_{a]} \c^\rho_i + 4 
(\G_{ba}\G_{c})_{\a\rho} V^c_{ij} \c^{\rho j}, \label{susygluon}\\
D_{\a i} \c^{\b}_j &=& {\delta_\a}^\b A_{ij} - \e_{ij} \frac{1}{4} 
{\G^{ab}}_\a{}^\b F_{ab}. \label{susygluino}
\eea
Here $A_{ij}$, symmetric in $i,j$, is a Lie-algebra valued auxiliary 
superfield which remains undetermined by the B.I. themselves. 
However, under the closure of the SUSY algebra \eref{algSUSY} on $\c^{\a 
i}$, i.e. applying one spinorial derivative to \eref{susygluino} and 
using on its r.h.s. \eref{susygluon}, one gets the consistency 
condition
\beq
D_{\a \,(i } A_{jk)} = - 4 \G^a_{\a\b} \, \c^\b_{(i} V_{a\,jk)}. 
\label{31}
\eeq
This is the constraint which will determine $A_{ij}$ almost uniquely 
once a $V_{a\,ij}$ satisfying \eref{18} has been found.

The constraints for the $H$-B.I.'s 
\beq
\label{idH}
d H = c_1 tr F^2 ,
\eeq

\nin
are also standard
\bea
H_{\a i \b j \g k} &=& 0, \\
H_{a \a i \b j} &=& - 2 \e_{ij} \G_{a\,\a\b} \phi ,\\
H_{ab \b i} &=& - {\G_{ab\,\b}}^\rho \l_{\rho i}
\eea

\nin
where $\phi$ is the dilaton superfield and $\l_{\a i} \equiv D_{\a i} 
\f$ is the gravitello.
\eref{idH} implies also the supersymmetry transformations 
\bea
D_{\a i}\l_{\b j} &=& 4 \phi \G_{c \a \b} V^c_{ij} + 2 c_1 \G_{c\a\b} 
\c_i \G^c \c_j + \e_{ij} (\G^a_{\a\b} D_a \phi - \frac{1}{12} 
H^{+}_{\a\b}). \\
D_{\a i} H_{abc} &=& -3 \G_{[ab\,\a}{}^\rho D_{c]} \l_{\rho i} - 3  
\G_{[ab\,\sigma}{}^\rho T_{c]\a}{}^\s \l_{\rho i} - 3 \G^s {}_{[a\, 
\a}{}^\rho  T^-_{bc]s} \l_{\rho i}+  \nonumber\\
&+& 6 \f \G_{[a \,\a\s} 
T_{bc]}{}^\s_i + 8 c_1 \G_{ca\,\a\b} tr(\c^b_i F_{bc]}),
\eea

\nin
where $H^+_{\a\b} \equiv (\G^{abc})_{\a\b} H_{abc}$, and the relation  
\beq
T^-_{abc} = \frac{1}{\f} H^-_{abc} +  \frac{c_1}{\f} tr(\c_i \G_{abc} 
\c^i),
\eeq

\nin
which leads to the identification of $T_{abc}^-$ as the $B_{ab}^-$ 
curvature. 

In all the cases we considered in this paper
the closure of the SUSY algebra on $\l_{\a i}$ 
is automatic once \eref{18} and \eref{31} are satisfied.

It is useful to notice that 
the consistency condition \eref{18} is equivalent to the existence
of a tensor $X_{k\a}$ such that 
\beq
D_{\a i} V^a_{jk} =\e_{i(j}X_{k)\a}.   
\eeq

\subsection{Consistency conditions}

\label{condizconsist}

>From the solution of the B.I.'s we found two auxiliary fields which 
permit to couple the various multiplets. Thus, $V_{a\,ij}$ and $A_{ij}$ 
represent the only freedom left (a part from $\omega_{\a\, i,jk}$).
We saw that these fields are restricted by the consistency conditions
\bea
\label{cond1}
D_{\a (i} V^a_{jk)} &=& 0,\\
\label{cond2}
D_{\a (i} A_{jk)} &=& - 4 \G^a_{\a\b} \c^\b_{(i} V_{a\ jk)},
\eea

\nin
for which we will give now the solution which leads to the minimally 
coupled SUGRA-TENSOR-YM theory. We make the Ansatz 
\bea
\label{V}
V^a_{ij} &=& {\g}(\f) \l_i \G^{a} \l_j + {\delta}(\f) \c_i 
\G^a \c_j, \\
\label{A}
A_{ij} &=& \b(\f) \l_{(i}\c_{j)},
\eea

\nin
where $\g$, $\delta$ and $\b$ are functions of $\f$ which have to be 
determined from \eref{cond1}-\eref{cond2}.

Inserting \eref{V} and \eref{A} in \eref{cond1} and \eref{cond2} we find 
that the consistency conditions are identically satisfied if and only 
if $\g$, $\delta$ and $\b$ satisfy the following set of coupled 
differential equations (where $\g^\prime  = \frac{d \g}{d \f}$ 
etc.):
\beq
\left\{
\bet{l}
$\g^\prime  - 32 \phi \g^2 = 0$, \\
$\delta^\prime  + \delta  \b  = 0$, \\
$c_1 \b  + 2 \delta  + 2 \phi \delta  \b  = 
0$, \\
$\b^\prime  - \b^2 - 16 \g  (\phi \b  + 1) = 
0$.
\eet
\right.
\eeq

\nin
The general solution of this set of equation is parametrized by a 
real parameter $k$ and is given by
\bea
\g (\phi) &=& -\frac{1}{k + 16 \phi^2} \label{35}\\
\b (\phi) &=& \frac{4}{\sqrt{16\phi^2 + k}} \\
\delta (\phi) &=& - \frac{c_1}{2\phi + 2\sqrt{\phi^2 
+\frac{k}{16}}} .\label{37}
\eea

The result \eref{35}-\eref{37} with the corresponding relations 
\eref{V}-\eref{A} for $V_{a\,ij}$ and $A_{ij}$ generalizes the 
solutions for the B.I.'s \eref{IbT}-\eref{IbH}  (with 
$c_2 = 0$) which one finds in the literature \cite{NG}. These solutions
can be obtained from \eref{35}--\eref{37} in the following limits.
For $k \to 0$ one gets
\beq
\g (\phi) = -\frac{1}{16 \phi^2}, \quad
\b (\phi) = \frac{1}{\phi}, \quad
\delta (\phi) = - \frac{c_1}{4\phi},
\eeq

\nin
while for $k\to \infty$ one obtains $\g = \b = \delta = 0$. 
The corresponding expressions for the auxiliary fields are
\bea
 k \to 0,& \displaystyle A_{ij} = \frac{1}{\f} \l_{(i}\c_{j)} , & V^a_{ij} = 
-\frac{1}{16\f^2} \l_i \G^a \l_j - \frac{c_1}{4\f} \c_i \G^a \c_j; 
\label{39}\\
k \to \infty , & A_{ij} = 0, & V^a_{ij}  = 0 . \label{40}
\eea

Equation \eref{39} reproduces the standard minimally coupled 
SUGRA-TENSOR-YM system while \eref{40} gives rise to the decoupled 
pure SUGRA and to the coupled  (rigid) TENSOR-YM theory \cite{BSS}.

At present only the choice \eref{39} gives rise to equations of 
motion which can be derived from an action \cite{compon}. 

The choice 
\eref{40} leads to equations of motion which describe on one hand a 
free YM theory and on the other hand a tensor multiplet which is 
coupled, through the Chern-Simons term, to the YM multiplet.
For this system the authors in \cite{BSS} were able to derive an 
action which describes, however, not only the propagation of the 
TENSOR-YM system, but also the propagation of a spurious additional 
tensor multiplet. 

A peculiarity of the tensor multiplet is the presence
of a self-dual boson.
Manifestly covariant actions for self-dual bosons
have been constructed only recently \cite{PST}; it does, 
however, not seem that the form of these actions can be generalized to 
describe the TENSOR-YM system, the principal problem being that the YM 
equations of motion are free while the tensor equations of motion are 
not. 

It is, however, possible to write a manifestly Lorentz-covariant 
and supersymmetric action for the free tensor multiplet 
and for the free supergravity multiplet
\cite{DLT}.

For a  generic $k$ one obtains equations of motion, for a coupled 
SUGRA-TENSOR-YM system, which close among them under SUSY (just like  
for the TENSOR-YM system), but at present it is not clear to us if 
this system admits an action. This point deserves further investigation. 
Notice, however, that the limiting cases \eref{39} and \eref{40} are 
the unique cases which preserve scale-invariance.

Notice also that in this section we showed that
the TENSOR-YM system can be obtained as a limiting case $(k\to 0)$, 
from a coupled 
SUGRA-TENSOR-YM system, in which SUGRA decouples. This presents,
in particular, the limiting procedure missed in \cite{BSS}.

Once one has an explicit solution for the consistency conditions,
it remains to derive the equations of motion for the physical fields. 
Since for our purposes  we 
do not need them we will not give them here explicitly. While 
Einstein's equation and the equations
for the fermions have to be derived always using standard superspace 
techniques the equations of motion for the abelian forms will be a
byproduct of our duality technique, which we illustrate  in the next
section.

\section{Duality}

In supergravity theories, all the 
bosonic fields, apart from the metric, are 
described through forms. We described the dilaton $\f$ with a 0-form, 
the gluon $A$ with the gauge connection 1-form and the graviphoton $B$
with a two-form, the curvatures associated to these potentials being  $F = 
dA + AA$, $H = d B + c_1\omega_{3YM}$ and, implicitly, we used also 
the one--form curvature for the dilaton $V = d\f$.

The number of 
bosonic physical degrees of freedom described by a $p$-form potential in 
$D$ dimensions is given by
\beq
\label{formule}
\left(\bet{c} $D-2$ \\ $p$ \eet \right) = \frac{(D-2)!}{p! 
(D-p-2)!}.
\eeq

\nin
which is manifestly invariant under $p \to (D-p-2)$. 
Therefore, as it is well known, at the kinematical level a
$p$-form potential, with a $(p+1)$-form as curvature, carries the same 
degrees of 
freedom as a $(D-p-2)$-form potential with a $(D-p-1)$-form as 
curvature, the two curvatures being related by Hodge duality.

The problem we address here is if this duality 
is compatible (and to which extent) with supersymmetry. 
In the superspace language the solution of this problem 
(see \cite{LT}) amounts to determine 
supercurvatures $\tF_4$, $\tH_3$ and $\tV_5$, associated respectively 
to $F_2$, $H_3$ and $V_1$, in such a way that their bosonic components 
are Hodge-dual to the bosonic components of $F_2$, $H_3$ and $V_1$ 
respectively, 
and that $\tF_4$, $\tH_3$ and $\tV_5$ satisfy B.I.'s  whose bosonic 
components are the equations of motion for the bosonic 
potentials $A_1$, $B_2$ and $\f_0$ we started with.
Moreover, and most importantly, these new B.I.'s have to be "true" B.I.'s 
in superspace
in the sense that they must allow to introduce new superpotentials
$\tA_3$, $\tB_2$ and $\tf_4$ which, in turn, can be interpreted as superspace 
duals of $A_1$, $B_2$ and $\f_0$. The original B.I.'s for $A_1$, $B_2$ and $\f_0$ 
can then be read as equations of motion for $\tA_3$, $\tB_2$ and 
$\tf_4$. 

In this section we will implement this program for the 
minimal SUGRA--TENSOR--SUPER MAXWELL--system (abelian gauge fields) and 
we will present its generalizations in the next section. The restriction to 
abelian gauge fields is necessary because in that case the gluons are 
decoupled from the gluinos, while in the non-abelian case the gluino 
current, appearing at the r.h.s. of the YM--equation of motion is 
"non-topological", in the sense that off--shell it can not be expressed 
as the differential of any local form. This prevents one from introducing 
dual non--abelian potentials $\tA_3$, see below. Therefore we 
restrict ourself to the gauge group $G = [U(1)]^N$ and $F \equiv 
(F_1, \ldots, F_N)$, where $F_i = d A_i$.

We recall that the standard B.I.'s are
\bea
\label{iid1}
dF &=& 0 \\
d H &=& c_1 trF^2 \label{iid2}\\
\label{iid3}
d V &=& 0.
\eea
Our Ansatz for the B.I.'s of the dual superforms $\tF, \tH$ and $\tV$ is 
the following (see also \cite{LT} for the ten dimensional case)
\bea
\label{id1}
d \tF &=&  F\tH \\
d \tH &=& 0 \label{id2}\\
\label{id3}
d \tV &=& c_1 tr( F \tF) + H\tH.
\eea
Notice first of all that this Ansatz is self-consistent in that, 
thanks to \eref{iid1}-\eref{iid3} the right hand sides of
\eref{id1}-\eref{id3} are closed forms.
Moreover, it is easy to see that the r.h.s. are also exact: it is this 
non trivial fact which will allow us, according to our above requirement,
to introduce dual super potentials in section 4.1.

We will now give a constructive proof of \eref{id1}-\eref{id3}
specifying the components of $\tH$, $\tF$ and $\tV$ which 
satisfy them identically. We define:
\bea
\label{conH1}
\tH_{\a i\b j\g k} &=& 0, \\
\tH_{a \a i \b j} &=& \frac{2}{\phi} \e_{ij} (\G_a)_{\a\b}, \label{conH2}\\
\tH_{ab \a i} &=& -\frac{1}{\phi^2} (\G_{ab})_\a{}^\b \l_{\b i}, \\
\tH_{abc} &=& \frac{1}{\phi^2} \frac{1}{3!} \e_{abcdef}\left(H^{def} - 
\frac{1}{4\phi} \l_i \G^{def}\l^i + c_1 tr(\c^i \G^{def} \c_i)\right),
\label{conH3}
\eea

\nin
for $\tH$,
\bea
\label{conF1}
\tF_{(p,4-p)}  &=& 0 \qquad p \leq 2,\\
\label{conF2}
\tF_{abc \a i} &=& -\frac{2}{\f} (\G_{abc})_{\a\b} \c^\b_i,\\
\tF_{sabc} &=& \frac{1}{ 2! \phi} \e_{sabcef}\left( F^{ef} +
\frac{1}{\phi} \c^k\G^{ef}\l_k \right),
\label{conF3}
\eea

\nin
for $\tF$ and
\bea
\label{conV1}
\tV_{(p,5-p)} &=& 0 \qquad p \leq 3,\\
\label{conV2}
\tV_{abcd\a i} &=& \frac{2}{\f} {\G_{a_1\ldots a_4\,\a}}^\rho \l_{\rho 
i},\\
\tV_{a_1 \ldots a_5} &=& - \frac{1}{\phi} \e_{a_1 \ldots a_6} \left( 
D^{a_6} \f - \frac{1}{4\phi} \l_k \G^{a_6} \l^k + 2 c_1 tr(\c_k \G^{a_6} 
\c^k)\right), \label{conVlast}
\eea
for $\tV$.
The proof that, with these definitions, the B.I's \eref{id1}-\eref{id3} are 
indeed satisfied goes as follows. First one proves \eref{id2}. Defining 
the four--superform $K_4 \equiv d\tH$ it is easy to show that $K_{(0,4)} = 
\ldots = K_{(3,2)} = 0$. At this point one uses the following 

\vskip0.2truecm
\nin
{\bf Lemma.} \ If a $p$-superform $K_p$, with $3 \leq p \leq 
6$, satisfies
\begin{enumerate}
\item $d K_p = 0$,
\item $K_{(0,p)} = \ldots = K_{(p-2,2)} = 0,$
\end{enumerate}
\vskip0.2truecm

\nin
then $K_p = 0$ identically, (see \cite{LT,Anom}).

\medskip

Since $d \tilde{K}_4 = 0$ the lemma implies now that $\tilde{K}_4=0$ and
\eref{id2} holds. If we define now the five-superform $K_5 \equiv d 
\tF - F \tH$, thanks to \eref{id2} and \eref{iid1} we have $d K_5 = 
0$ and it is straightforward to show that $K_{(0,5)} = \ldots = 
K_{(3,2)} = 0$. 
The lemma implies then that $K_5 = 0$ and also \eref{id1} holds. Finally 
\eref{iid1}, \eref{iid2}, \eref{id2} and \eref{id1} ensure now that 
$K_6 \equiv d\tV - c_1 tr(F \tF) - H \tH$ satisfies $d K_6 = 0$ and a 
direct calculation gives $K_{(0,6)} = \ldots = K_{(4,2)}= 0$. The 
lemma implies then also \eref{id3}. 

What we proved is essentially 
that \eref{iid1}-\eref{iid3} imply \eref{id1}-\eref{id3}.
But, since the purely bosonic components of the dual supercurvatures 
are defined essentially as the Hodge-duals of the basic bosonic 
curvatures (see equations \eref{conH3}, \eref{conF3} and 
\eref{conVlast}), the purely bosonic components of 
\eref{id1}-\eref{id3} correspond to the equations of motion for 
$A_1$, $B_2$ and $\f_0$ respectively.
This is clearly no surprise  since we know that the B.I.'s 
\eref{iid1}-\eref{iid3}, under a suitable choice of constraints, set 
the theory on-shell. Therefore, our procedure for dualizing the 
abelian connections can also be regarded as an alternative way for 
deriving their equations of motion.

\subsection{The dual superconnections}

The identities \eref{id1}-\eref{id3} allow now the 
introduction of dual potentials $\tB_2$, $\tA_3$, $\tf_4$ according to 
\bea
\tH &=& d \tB, \label{111}\\
\tF &=& d \tA + \tB F,\label{222}\\
\tV &=& d \tf + H \tB + c_1 F \tA. \label{333}
\eea

The possibility of describing the three-form $H$ in terms of a $B_2$ 
potential or a dual $\tB_2$ potential in compatibility with 
supersymmetry is of course known from a long timeÊ\cite{NG}.

The existence of dual $\tA_3$ potentials for Maxwell fields has been 
conjectured by Schwarz and Sen \cite{SSn} in ten dimensions 
(in which case they become 
7-superforms) in relation with the existence of a manifestly $SL(2, 
\hbox{\msbm R} )_S $ invariant form for the heterotic string effective 
action compactified down to four dimensions. These seven-superform 
gauge fields have actually been constructed in ten-dimensional 
superspace \cite{LT} so that the existence of their six dimensional 
counterpart (three-superforms) in \eref{222} constitutes actually no 
surprise: they could also have been obtained by compactifying $N = 
1$, $D = 10$ SUGRA-MAXWELL toroidally down to six dimensions and then 
truncating the resulting $N= 2$, $D = 6$ theory to an $N = 1$ 
supersymmetric SUGRA-MAXWELL-TENSOR theory. The construction we gave 
here, however, is direct and independent of the details of any 
compactification scheme.

The significance of the supersymmetric dualization of the dilaton is, 
on the other hand, less clear. To our present knowledge the dual 
potential $\tf_4$ appearing in \eref{333} did not have any direct 
application but it may be, for instance, that in some supersymmetric 
three-brane $\s$-model it can be coupled "naturally" to the brane 
through its pull-back on the four-dimensional brane world-volume.

There are, however, important differences between the features of 
$\tB_2$ on one hand, and the features of $\tA_3$ and $\tf_5$ on the other. 
If one uses $\tB_2$ to describe the graviphoton and $A_1$ and $\f_0$ 
to describe respectively the Maxwell fields and the dilaton (the 
standard dual $N = 1$, $ D = 6$ SUGRA \cite{NG}) then in all the 
equations of motion the potentials appear only through their 
curvatures $\tH_3$, $F_2$ and $V_1$, and the theory can be described in 
terms of an action involving $\tB_2$, $A_1$ and $\f_0$.
 
If, on the other 
hand,  one tries to describe the theory in terms of $\tB_2$, $\tA_3$ and 
$\f_0$, then one is faced with the problem of eliminating from all the 
equations of motion $A_1$ in favour of $\tA_3$; in particular in 
 \eref{222} the curvature $F$, which has now to be viewed as the dual of 
$\tF$, appears on the r.h.s. so that this equation determines $\tF$ 
in terms of $\tA_3$ only implicitly and it is not possible to write an 
action, at least in closed form, in which $A_1$ has been replaced by $ 
\tA_3$. 
The situation is even worse when one tries to use $\tf_4$ instead 
of $\f_0$ using \eref{333}: in this case the dilaton $\f_0$ appears at the 
r.h.s. of \eref{333} explicitly, i.e. not through its 
curvature $V = d\f_0$, and it is not possible, not even implicitly, to 
eliminate $\f_0$ completely from the game in favour of $\tf_4$. 
Nevertheless, at the level of equations of motion,  
\eref{111}-\eref{333} are perfectly consistent with 
\eref{id1}-\eref{id3}. 
For example, gauge invariance for $\tB$ in \eref{222}, $\tB \to \tB + 
dC$, is saved, upon using $d F = 0$, by imposing $\tA \to \tA - CF$. 

\section{Introducing the Hypermultiplets}

Supersymmetry in six-dimensions allows the existence of matter 
fields, i.e. of hypermultiplets.
In this section we generalize our results concerning duality to the 
presence of these fields for the "ungauged" case, i.e. when they are 
charge-less with respect to the abelian Maxwell fields.

As shown in \cite{N,W1} and \cite{Materia}
self-interacting hypermultiplets in a $N= 1, D = 6$ theory live on a 
quaternionic K\"ahler manifold. 
For simplicity we will use the particular coset manifold $USp(n_H, 
1) / USp(n_H) \otimes USp(1)$ where $n_H$ is the number of hypermultiplets
\cite{N,NG}.
In what follows the indices $I,J = 1, \ldots, 4n_H$ are used for the 
hyperscalars which parametrize the quaternionic manifold,  
$i,j = 1,2$ are $USp(1)$ indices and $X,Y,Z = 1, 
\ldots, 2 n_H$ label the fundamental representation of $USp(n_H)$.
$g_{IJ}(\f)$ indicates the quaternionic manifold metric tensor, 
$\omega_{Ii}{}^j (\vf)$ and $\omega_{IX}{}^Y (\vf)$ are the 
$USp(1)$ and $USp(n_H)$ connections and $e^I_{iZ}$ are 
the coset vielbeins. For completeness
we remember also the relations obeyed by the vielbeins:
\bea
g_{IJ} \ e^I_{i X} \ e^J_{j Z} &=& \e_{ij} \  \e_{XZ}, \\
e^I_{i Z} \  e^{J j Z} + e^J_{i Z} \ e^{I j Z} &=& g^{IJ} \ {\delta_i}^{j}, \\
e^I_{i Y} \ e^{J i Z} + e^J_{i Y} \ e^{I i Z} &=& \frac{1}{n_H} \ g^{IJ} 
{\delta_Y}^{Z},
\eea
where $\e_{ij}$ and $\e_{XY}$ are the invariant tensors of $USp(1)$ 
and $USp(n_H)$ respectively.

As we saw in section three
the only freedom left in the solution of the torsion B.I.'s were the 
three spinors $\omega_{\a i, jk}$  in $T_{\a i \b j}{}^{\g k}$.
A natural choice for the introduction of hypermultiplets \cite{NG} 
amounts to the identification of $\omega_{\a i, jk}$ with the $USp(1)$ 
connection. This connection, which realizes local $USp(1)$ covariance,
is a
function of the hypermultiplet superscalars $\vf^I (Z)$
\beq
\omega_{ij}(\vf) = d \vf^I \omega_{I\,ij}(\vf),
\eeq
its pull-back on the cotangent bundle basis of superspace being given by
\beq
\omega_{ij} = E^{A} D_A \vf^I \omega_{I\,ij} \equiv E^{\a 
k} \omega_{\a k, ij} + E^a \omega_{a\,ij}. 
\eeq
The new torsion constraint would  now read  
\beq
T_{\b j \g k}{}^{\a i} =  \delta_\b^\a \omega_{\g k, j}{}^i + (\b j 
\leftrightarrow \g k) \label{star}
\eeq
and one should solve the torsion B.I.'s with this new dynamical 
constraint \cite{NG}.

It is, however, more convenient to proceed in a slightly 
different, but equivalent, way. Instead of imposing the new constraint 
\eref{star} on the torsion, which would then turn out to transform 
under $USp(1)$ as 
a connection and not as vector, we define a new torsion 
which is $USp(1)$ covariant and satisfies, moreover, the old 
constraint, $T_{\a i \b j}{}^{\g k} = 0$:
\bea
T^a &\equiv& d E^a + E^b {\Omega_b}^a, \label{def0}\\
T^{\a i} &\equiv& \D E^{\a i}=d E^{\a i} + E^{\b i} {\Omega_\b}^\a + E^{\a j} 
{\omega_j}^i.\label{def}
\eea
It is also convenient to define a $USp(1)$ and $USp(n_H)$
covariant derivative $\D_A$ through
\bea
\D_A &=& D_A + D_A \vf^I \; {\omega_{I\#}}^\#
\eea
where $\omega_I$ equals $\omega_{Ii}{}^{j}$ or $\omega_{IX}{}^{Y}$ 
or both of them according to the tensor on which it acts.

With respect to the standard procedure 
our procedure has the advantage of being manifestly 
$USp(1)$ and $USp(n_H)$ covariant in that the connections appear
in the B.I.'s and their solutions only through the covariant derivatives.

Closure of the SUSY algebra on the hypermultiplets entails 
\bea
D_{\a i} \vf^I &=& e^I_{iZ}\psi^Z_\a , \\
\D_{\a i} \psi^Z_\b &=& 2 \; e^Z_{iI} \; \G^a_{\a\b} \; D_a \vf^I,
\label{512}
\eea
where the hyperfermions $\psi^Z_\a$ are the supersymmetric partners of 
$\vf^I$. This implies in particular that  
$\omega_{\a i, j}{}^k=D_{\a i}\vf^I\omega_{Ij}{}^k=
e^I_{iZ}\psi^Z_\a\omega_{Ij}{}^k$.
\nin

Once all derivatives have been covariantized,
the introduction of the hypermultiplets in our  
formalism leads essentially only to a change of the torsion B.I.'s
themselves. The definitions \eref{def0}, \eref{def} give, in fact, rise to the 
new B.I.'s
\bea
D T^a &=& E^b{ R_b}^a, \\
\D T^{\a i} &=& E^{\b i}{R_\b}^\a + E^{\a j} {\R_j}^i ,\label{514}
\eea

\nin
where $\R_j{}^i = d \omega_j{}^i + \omega_j{}^k \omega_k{}^i = 1/2 \, d 
\vf^I\, d \vf^J \, \R_{JI\, j}{}^i$ is the $USp(1)$ curvature.
As is well known SUSY requires the quaternionic manifold to be 
maximally symmetric, that is, in our conventions
\beq
\R_{IJ ij} =  2 ( e_{I i Z} e_{J j}{}^Z - e_{JiZ} e_{Ij}{}^Z).
\label{515}
\eeq
The new torsion B.I.'s can now again be consistently solved with the 
basic constraints
\bea
 T^a_{\a i \b j} &=& -2 \e_{ij} \G^a_{\a\b}, \\
T_{\a i \b j}{}^{\g k} = &0& = T_{a\a i}{}^b.
\eea
The differences with respect to the case without hypermultiplets arise
now primarily from the new term on the r.h.s. of \eref{514}, i.e. 
$E^{\a j} \R_j{}^i$; all new terms are however {\it manifestly} covariant.

Taking \eref{512} and \eref{515} into account one finds for the 
torsion and $SO(1,5)$--curvature:
\bea 
T_{abc} &=& T^-_{abc} - \frac{1}{4} \psi^+_{abc}, \\
T_{a \a i}{}^{\b}_j &=& 3 {\delta_\a}^\b V_{a ij} + \G_{ac\,\a}{}^\b 
V^c_{ij} - \frac{1}{4} \e_{ij} \G^{bc}{}_\a{}^\b \left(T^-_{abc} -  
\frac{1}{4} \psi^+_{abc}\right), \\
R_{\a i \b j ab} &=& -4 \G_{abc\, \a\b} V^c_{ij}.
\eea
where the self-dual tensor $\psi^+_{abc}$ is defined by
\beq
\psi^+_{abc} \equiv \psi_Y \G_{abc} \psi^Y.
\eeq
For the YM-- and tensor--multiplets the relevant new SUSY 
transformations are now
\bea
\D_{\a i} \c^{\b}_j &=& {\delta_\a}^\b A_{ij} - \e_{ij} \frac{1}{4} 
{\G^{ab}}_\a{}^\b F_{ab}, \\
\D_{\a i} F_{ab} &=& 4 \G_{[b \ \a\rho} \D_{a]} \c^\rho_i + 4 
(\G_{ba}\G_{c})_{\a\rho} V^c_{ij} \c^{\rho j} + \frac{1}{4} 
{\G^{cd}}_{[b\,\a\rho} \psi^+_{a]cd} \c^\rho_i, \\
\D_{\a i}\l_{\b j} &=& 4 \phi \G_{c \a \b} V^c_{ij} + c_1 \G_{c\a\b} 
\c_i \G^c \c_j + \e_{ij} \G^a_{\a\b} D_a \phi - \frac{1}{12} \e_{ij}
\G^{abc}_{\a\b} \left(H^{+}_{abc} + \frac{\f}{4} \psi^+_{abc}\right).
\eea

\nin
while in particular the relation between $H^-_{abc}$ and $T^-_{abc}$ 
is unchanged. The consistency conditions remain formally the same
\bea
\D_{\a (i} V^a_{jk)} &=& 0,\\
\D_{\a (i} A_{jk)} &=& - 4 \G^a_{\a\b} \c^\b_{(i} V_{a\ jk)},
\eea
\nin
with the simple change $D_{\a i} \to \D_{\a i}$ with respect to 
equations \eref{cond1} and \eref{cond2}, and they 
are again solved by equations \eref{39}.
\medskip

Now it is  easy to solve the B.I.'s for the dual 
superforms $\tH$, $\tF$ and $\tV$. Indeed, these dual supercurvatures 
are again given by \eref{conH1}-\eref{conVlast} with the 
{\it unique} difference that now 
\beq
\tH_{abc} = \frac{1}{\phi^2} \frac{1}{3!} \e_{abcdef}\left(H^{def} - 
\frac{1}{4\phi} \l_i \G^{def}\l^i - \frac{\f}{2} \psi^{+def} 
+ c_1 tr(\c^i \G^{def} \c_i)\right).
\eeq

\section{Non minimal models: the Lorentz Chern-Simons form}

The dimensions $D = 2,6,10$ are special in  many respects. They allow 
for example, in a space-time with Minkowskian signature, the 
existence of chiral bosons with self-dual curvatures, 
respectively one-, three- and five-forms.

They are also the unique dimensions below eleven
which are potentially plagued by Lorentz 
anomalies. In all physically significant theories, however, the total 
anomaly polynomial factorizes and the anomaly can be cancelled via the 
Green-Schwarz mechanism \cite{GSM}. The essential ingredients are in 
each case a modified B.I. for a generalized three-form curvature, in 
ordinary bosonic  space,
\beq
d H = c_1 tr F^2 + c_2 tr R^2,
\label{sstar}
\eeq

\nin
(a classical effect) and the subtraction from the effective action of 
a local counterterm, proportional to $B_2$ (a quantum effect).
Both ingredients, however, do not directly respect supersymmetry. Since 
\eref{sstar} is a local and classical relation its 
supersymmetrization can be achieved by solving the corresponding B.I. 
in {\it superspace}.
The analogous problem in $N=1,D=10$ supergravity 
has been solved in \cite{BPT,catena} with the aid of 
the Bonora-Pasti-Tonin [BPT] theorem. The aim of the present section 
is to "extend" this theorem to six-dimensional supergravity and to 
solve \eref{sstar} for $c_2\neq 0$ in superspace. 
At the level of the action the Lorentz--Chern--Simons form 
gives rise to non-minimal couplings of the form 
$(R_{abcd})^2$, together with their supersymmetric 
completion. 

Having solved \eref{sstar}, in the next section we will 
discuss the existence of the dual superpotentials $\tB$, $\tA$ and 
$\tf$ in the resulting non-minimal $N = 1$, $D = 6$ supergravity.

For the sake of simplicity we turn back to the hypermultiplet free 
model considered in section three. For convenience we repeat here the 
general parametrizations of the torsion and curvatures we obtained:
\bea
\label{T1}
T_{\a i \b j}{}^a &=& -2 \e_{ij} \G^a_{\a\b},\\
{T_{\a i a}}^b &=& 0 =  {T_{\a i \b j}}^{\g k}, \\
T_{abc}  &=& T^-_{abc}, \\
{T_{a \a i}}^{\b}_j &=& 3 {\delta_{\a}}^\b V_{a ij} + {\G_{ac \a}}^\b 
V^c_{ij} - \e_{ij} \frac{1}{4} {\G^{bc}}_\a{}^\b T^-_{abc},\\
{R_{\a i\b j ab}} &=& -4 \G_{abc\,\a\b} V^c_{ij}, \label{66}\\
{R_{a\a i b c}} &=&  - 2 \G_{a\,\a\b} {T^\b_{bc i}} + 
{\G_{[ab\,\a}}^\b \left( \frac{2}{3} \e^{hk} {\G_{c]s \b}}^\rho D_{\rho h} 
 V_{ki}^s - \frac{4}{3} \e^{hk} D_{\b h} V_{c] ki}\right),\\
D_{\a i}T^-_{abc}&=& 6(\G_{[a})_{\a\b}T^{\b}_{bc]i} - 2 \G_{[ab}{}_\a 
{}^\b \left( \frac{2}{3} \e^{hk} {\G_{c]s \b}}^\rho D_{\rho h} 
 V_{ki}^s - \frac{4}{3} \e^{hk} D_{\b h} V_{c] ki} \right),\\
D_{\a i} V^a_{jk} &=& \e_{i(j} X_{k)\a}.
\label{Tend}
\eea
We included here in the last equation also the consistency condition 
for the unknown auxiliary field $V_{ij}^a$; at the end we will 
obtain an (implicit) expression for this field in terms of the physical
fields, see \eref{646} below, and one has to check if
it satisfies \eref{Tend}.  

\vskip0.2truecm
\paragraph{BPT Theorem.} The parametrizations \eref{T1}-\eref{Tend} 
allow for the decomposition
\beq
\label{32}
tr R^2 = d X + K,
\eeq
\nin
where $X$ and $K$ are gauge invariant three and four-superforms 
respectively and
\beq
\label{33}
K_{(0,4)} = 0 = K_{(1,3)}.
\eeq
\vskip0.2truecm

\nin
A simple consequence of $d trR^2 = 0$ is that $d K = 0$.
Before proving the theorem we remark that if we define
\beq
\hat{H} = H - c_2 X
\eeq

\nin
the $H$-B.I. becomes
\beq
d \hat{H} = c_1 tr F^2 + c_2 K.
\eeq
This identity can now be solved imposing on $\hat{H}$ the "old" constraints
\bea
\hat{H}_{\a i\b j\g k} &=& 0, \\
\hat{H}_{a\a i \b j} &=& - 2\phi \e_{ij} (\G_a)_{\a\b} , \\
\hat{H}_{\a i ab} &=& - \G_{ab\, \a}{}^\b \l_{\b i} 
\eea
since the four-form $K$ has the same structure as $tr F^2$: 
$d\ trF^2=0$, $(tr F^2)_{(0,4)}=(tr F^2)_{(1,3)}=0$.

For the proof of the theorem we adopt the techniques used in \cite{BPT}. 
We write the superspace differential $d$ (when it acts on Lorentz {\it 
invariant} forms) as a sum 
of operators
\beq
d = \ol{d} + \ol{D} + T + \tau
\label{decompo}
\eeq

\nin
each of which sends a $(p,q)$-superform to a 
$(p^\prime,q^\prime)$-superform. Defining their degree as the 
difference $(p^\prime-p, q^\prime - q)$, the operators 
$\ol{d}$, $\ol{D}$, $T$, $\tau$ have respectively degree $(1,0)$, $(0,1)$, 
$(-1,2)$ and $(2,1)$.

In more detail: $\ol{d} = E^a D_a + T_{(1,0)}$,  $\ol{D} = E^{\a i} 
D_{\a i} + T_{(0,1)}$, $T = T_{(-1,2)}$, $\tau = T_{(2,-1)}$; where 
$D_a$ and $D_{\a i}$ 
are the ordinary covariant derivatives acting only on the components, 
while $T_{(r,s)}$ acts only on the vielbeins as follows:
\bea
T_{(-1,2)} E^a = \e_{ij} \G^a_{\a \b} E^{\a i} E^{\b j}, && T_{(-1,2)} 
E^{\a i} = 0, \\
T_{(2,-1)} E^a = 0, && T_{(2,-1)} E^{\a i} = \frac{1}{2} E^b E^c 
{T_{cb}}^{\a i}, \\
T_{(1,0)} E^a =  \frac{1}{2} E^b E^c {T_{cb}}^a, && T_{(1,0)} 
E^{\a k} = E^{\b i} E^b {T_{b\b i}}^{\a k}, \\
T_{(0,1)} E^a = 0, && T_{(0,1)} E^{\a i} = 0.
\eea
Due to  $d^2 = 0$ these operators satisfy the following anticommutation 
rules:
\bea
\label{antprop}
T^2 = 0 \nonumber \\
\ol{D} T + T \ol{D} = 0 \nonumber \\
\ol{D}^2 + \ol{d} T + T \ol{d} = 0 \nonumber\\
\ol{dD} + \ol{Dd} + T \tau + \tau T = 0 \\
\ol{d}^2 + \ol{D} \tau + \tau \ol{D} = 0 \nonumber \\
\ol{d}\tau + \tau \ol{d} = 0 \nonumber \\
\tau^2 = 0. 
\eea
A crucial ingredient of the proof is in particular the operator $T$ which, 
as direct consequence of the cyclic identity, is a coboundary operator
\beq
T^2 = 0.
\eeq
If we call $S_{(p,q)}$ the space of $(p,q)$-superforms, then $T$ maps 
$S_{(p,q)}$ to $S_{(p-1,q+2)}$.

Setting  $Q \equiv tr R^2$, our starting point is the identity
\beq
d Q = 0.
\eeq
Projecting it on the sectors $(p,5-p)$, we obtain
\bea
\label{316a}
T Q_{(0,4)} = 0, \\
\label{316b}
T Q_{(1,3)} + \ol{D} Q_{(0,4)} = 0, \\
\label{316c}
T Q_{(2,2)} + \ol{D} Q_{(1,3)} + \ol{d} Q_{(0,4)} = 0.
\eea

\nin
The identity \eref{316a} states that $Q_{(0,4)}$ is a $T$-cocycle;
the first step is to verify that  it is a trivial one. 
This means that there exists a $ X_{(1,2)} \in 
S_{(1,2)}$ such that
\beq
\label{317}
Q_{(0,4)} = T X_{(1,2)}.
\eeq
The explicit expression for $Q_{(0,4)}$ can be obtained from the 
expression of $R_{\a i\b j a}{}^b$ \eref{66}
\beq
Q_{(0,4)} = 4 E^{\delta l} E^{\g k} E^{\b j} E^{\a i} 
((\G_{abc})_{\a\b} (\G^{bas})_{\g\delta} V^c_{ij} V_{s\ kl})
\eeq

\nin
and one  easily finds that indeed
\beq
\label{621}
Q_{(0,4)} = T \ol{X}_{(1,2)}
\eeq

\nin
with
\beq
\ol{X}_{(1,2)} = {3\over 2}E^a E^{\a k} \G^b_{\a\b} E^{\b}_k  
V_{aij} V_b^{ij}.
\eeq

\nin
Clearly $X_{(1,2)}$ is defined only modulo a non-trivial $T$-cocycle 
$Y_{(1,2)} \in S_{(1,2)}$ so that we can set more in general 
\beq
\label{320}
X_{(1,2)} = \ol{X}_{(1,2)} + Y_{(1,2)}, \qquad T Y_{(1,2)} = 0.
\eeq
We neglect here trivial $T$-cocycles $\in S_{(1,2)}$
since they 
express simply the fact that $X$ is defined modulo trivial cocycles of 
$d$ itself which  amount to a redefinition of $B$.
The  general non-trivial cocycles of $T$ in $S_{(1,2)}$ are of the form
\beq
Y_{a\a i\b j} = (\G_{abc})_{\a\b} Y^{bc}_{ij}, \qquad Y^{bc}_{ij} =  
Y^{[bc]}_{(ij)},
\eeq

\nin
if we disregard terms of the form
\beq
\e_{ij} {\G_{a}}_{\a\b} \hat{\phi}
\eeq

\nin
which amount to a redefinition of  $\phi$.
On dimensional grounds there are only three terms which can 
contribute to $Y^{bc}_{ij}$:
\bea
\label{prima}
Y^{(1)}_{bc\, ij} &=& D_{[b} V_{c] ij} \\
\label{seconda}
Y^{(2)}_{bc\, ij} &=& V_{[b\,i}{}^k V_{c]\,jk}  \\
\label{terza}
Y^{(3)}_{bc\, ij} &=& T^-_{abc} V^a_{ij}.
\eea

\nin
Inserting now \eref{317} in \eref{316b}, we 
obtain
\beq
\label{TQc}
T(Q_{(1,3)} - \ol{D} X_{(1,2)}) = 0,
\eeq

\nin
which means that $(Q_{(1,3)} - \ol{D} X_{(1,2)}) \in S_{(1,3)}$ is a 
$T$-cocycle. It can be seen that for a generic $Y_{(1,2)}$ it is a 
non--trivial one.
An explicit computation reveals, in fact, that it becomes a trivial 
cocycle if and only if one chooses
\beq
\label{alpha}
Y^{bc}_{ij} = {Y^{(3)}}^{bc}_{ij} + \a \frac{1}{3} {Y^{(1)}}^{bc}_{ij}
 + (1-\a) {Y^{(2)}}^{bc}_{ij},
\eeq
where $\a$ is an arbitrary constant.

We will comment on the correct choice for the parameter $\a$ below. 
Here we would only like to remark the following. If $\a \neq 0$ 
then $Y_{(1,2)}$ would contain a term linear in the bosonic
derivative of the still undetermined auxiliary field $V^a_{ij}$. The 
$H$-B.I. would then give rise to an equation for this field, see 
\eref{646} below, of the kind
\beq
V^ a_{ij} = c \, \Box V^a_{ij} + \ldots, \label{str}
\eeq

\nin
where $c$ is a constant, proportional to $\a$, and the remaining terms 
are quadratic and of higher order in $V^a_{ij}$. This equation, which 
is the generalization of \eref{39} to $c_2\neq 0$,  would 
now propagate unphysical spurious degrees of freedom due to the 
appearance of the D'Alambertian. 
It can be seen that these degrees of freedom are actually 
unitary-violating poltergeists. The presence of such poltergeists in 
supergravity theories with Lorentz-Chern-Simons terms is indeed known 
from a long time. Their supersymmetric structure, their meaning and 
their possible elimination, has been discussed in detail in \cite{BPT}, 
to which we refer the reader for more details.

With the choice \eref{alpha}, $Q_{(1,3)} - \ol{D} X_{(1,2)}$ becomes a 
trivial $T$-cocycle and the equation 
\beq
Q_{(1,3)} = T X_{(2,1)} + \ol{D} X_{(1,2)}
\label{633}
\eeq

\nin
determines the three-form $X_{(2,1)}$ uniquely since in the sector 
$(2,1)$ there are no $T$- cocycles at all: 
\beq
T X_{(2,1)} = 0 \Leftrightarrow
X_{(2,1)} = 0.
\eeq
The explicit expression for $X_{(2,1)}$ will not be 
needed in what follows however.

 Inserting now 
\eref{317} and \eref{633} in \eref{316c} and using the anticommutation 
properties of $T,\ol{D},\ol{d}$ \eref{antprop} one gets 
\beq
T(Q_{(2,2)} - \ol{D} X_{(2,1)} - \ol{d} X_{(1,2)}) \equiv T W_{(2,2)} = 0.
\eeq

\nin
The non-trivial $T$-cocycles in the $(2,2)$ sector have the general 
structure
\beq
K_{(2,2)} = E^a E^b E^{\a i} E^{\b j} (- (\G_a)_{\a\g} (\G_b)_{\b\delta} 
L^{\g\delta}_{ij}),
\label{635}
\eeq

\nin
where $L^{\g\delta}_{ij} = L^{[\g\delta]}_{(ij)}$,
and therefore
\beq
W_{(2,2)} = T X_{(3,0)} + K_{(2,2)}.
\eeq

Again, the forms  $X_{(3,0)}$ and $K_{(2,2)}$, i.e. 
$L^{\g\delta}_{ij}$, are uniquely determined by the above formulae, 
but their explicit expressions would turn out to be rather 
complicated and we do not need them here. In conclusion, we got the last 
decomposition needed to prove the six-dimensional version of the BPT 
theorem:
\beq
Q_{(2,2)} = T X_{(3,0)} + \ol{D} X_{(2,1)} + \ol{d} X_{(1,2)} + 
K_{(2,2)}. \label{doubstar}
\eeq
Indeed, 
formulae \eref{317},\eref{633} and \eref{doubstar} combine now exactly 
to give \eref{32} if one defines
\beq
X = X_{(1,2)} + X_{(2,1)} + X_{(3,0)}.
\eeq
>From these formulae one sees also that $K_{(0,4)} = K_{(1,3)} = 0$ 
and that $K_{(2,2)}$ is given in \eref{635}. Notice that at this 
point the knowledge of $K_{(2,2)}$ and the identity $d K = 0$ 
determine the four-superform $K$ uniquely (use again the lemma). 
Notice also that  $K_{(2,2)}$, given in \eref{635}, has indeed the same 
structure as $(tr F^2)_{(2,2)}$.

\hfill {\bf Q.E.D.}

\subsection{Consistency conditions}

We did not really end our proof of the compatibility between the
Lorentz--Chern--Simons form and supersymmetry. We still have to show that the  
$H$-B.I. can be solved consistently and that the consistency 
conditions \eref{cond1}-\eref{cond2} are satisfied.
To this end, as anticipated, we define $\hat{H} = H - c_2 X$, whose B.I. 
reads $d \hat{H} = c_1 trF^2 + c_2K$, i.e.
\bea
(d \hat{H})_{(0,4)} &=& 0 \label{i37}\\
(d \hat{H})_{(1,3)} &=& 0 \label{i38}\\
(d \hat{H})_{(2,2)} &=& c_1 (trF^2)_{(2,2)} + c_2 K_{(2,2)} \label{i39}\\
(d \hat{H})_{(3,1)} &=& c_1 (trF^2)_{(3,1)} + c_2 K_{(3,1)} \\
(d \hat{H})_{(4,0)} &=& c_1 (trF^2)_{(4,0)} + c_2 K_{(4,0)} 
\eea

\nin
and, due to the lemma, it is sufficient  to solve \eref{i37}-\eref{i39}.
On $\hat{H}$ we impose the constraints
\bea
\hat{H}_{\a i\b j\g k} &=& 0, \\
\hat{H}_{a\a i \b j} &=& - 2\phi \e_{ij} (\G_a)_{\a\b},Ê\\
\hat{H}_{ab \a i} &=& - (\G_{ab})_\a{}^\b \l_{\b i},
\eea
which solve already \eref{i37}-\eref{i38}.
Equation \eref{i39} implies (and is solved) by the relations:
\bea
D_{\a i} \l_{\b j} &=&  4 \phi \G_{c \a \b} V^c_{ij} 
-2(c_1\c_{\a\b\, ij}+c_2L_{\a\b\, ij})
 + \e_{ij} (\G^a_{\a\b} D_a \phi - \frac{1}{12} 
\hat{H}^{+}_{\a\b}).\\
T^-_{abc} &=& \frac{1}{\f} \hat{H}^-_{abc} + \frac{c_1}{\f} tr(\c_i 
\G_{abc} \c^i),
\eea
where
\bea
\c_{\a\b\, ij}  &=& - \G^a_{\a\b} tr(\c_i \G_a \c_j), \\
L_{\a\b\, ij}  &=& - \G_{a\,\a\b} (\G^a_{\g\delta} L^{\g\delta}_{ij}).
\eea

We  remain finally only with the consistency conditions:
\bea
D_{\a (i} A_{jk)} &=& 4 \G^a_{\a\b} \c^\b_{(i} V_{a\,jk)},\label{firstc} \\
D_{\a (i} V^a_{jk)} &=& 0. \label{2ndcond}
\eea

\nin
Maintaining for $A_{ij}$ the definition in \eref{39},
the first condition tells us that $V^a_{ij}$ has to satisfy the 
equation
\beq
 V^a_{ij} = -\frac{1}{16\phi^2} \l_i \G^a \l_j - \frac{1}{16\f}\G^{a\,\a\b} 
 (c_1 \c_{\a\b ij}  + c_2 L_{\a\b ij}).
 \label{646}
\eeq
Let us stress that even if the super YM-fields are absent ($F=0$),
in which case the consistency condition \eref{firstc} becomes
trivial, this equation is implied anyway by the closure of the SUSY
algebra on $\l_{\a i}$.

This equation determines $V^a_{ij}$ only implicitly 
in that $L^{\g\delta}_{ij}$ is a complicated expression which involves
$V^a_{ij}$ itself in a non polynomial way (a part from the linear term
discussed in \eref{str}).

The second condition instead implies a new constraint on the $L$ 
supersymmetry transformation \footnote{This constraint is identically
satisfied by $\c_{\a\b ij}$}:
\beq
\label{conssst}
D_{\a}^{(i}L_{\b\g}^{jk)} = \frac{3}{\phi} \l_{[\a}^{(i}L_{\b\g]}^{jk)}.
\eeq

\nin
The identity $dK = 0$ ensures that $D_{\a}^{(i}L_{\b\g }^{jk)} $
is antisymmetric in $\a\b\g$
\beq
\label{650}
D_{\a}^{(i}L_{\b\g}^{ jk)} = D_{[\a}^{(i}L_{\b\g]}^{ jk)}
\eeq

\nin
which is the right structure required by \eref{conssst},
but we did not perform the long calculations needed to prove that 
$D_{[\a}^{(i}L_{\b\g]}^{ jk)}$ equals exactly $\frac{3}{\phi} 
\l_{[\a}^{(i}L_{\b\g]}^{jk)}$. 

These calculations would first of all require 
the explicit expression for $X_{(2,1)}$ which can be obtained
from \eref{633}; this 
could then be inserted in \eref{doubstar} to 
compute the explicit formula for $L^{\a\b}_{ij}$. The algebraic 
manipulation involved are straightforward, from a technical point of 
view, but very lengthy. So this is the missing point in our proof of 
the compatibility of the Lorentz-Chern-Simons form with SUSY.

Let us notice however that to satisfy \eref{650} we have still the 
freedom to choose the parameter $\a$ in \eref{alpha} arbitrarily. 
Indeed, since the poltergeists are present also in $N = 1$, $D =10$ 
SUGRA with a Lorentz-Chern-Simons term, and the present theory could be 
obtained from the ten-dimensional one upon compactification and 
truncation from $N = 2$ to $N = 1$ SUSY, we expect a non-vanishing $\a$.

Even if \eref{conssst} can not be satisfied for any $\a$, our 
procedure leaves another possibility open. 
In fact, as has been noted for the ten-dimensional case \cite{BPT}, 
the decomposition $trR^2 = dX + K$ is not unique. Suppose, in fact, 
that there exist non-exact three-superforms $Z$ satisfying
\beq
(dZ)_{(0,4)} = 0 = (dZ)_{(1,3)}.
\eeq

\nin
Then one can write
\beq
trR^2 = d \hat{X} + \hat{K},
\eeq

\nin
where now $\hat{X} = X + Z$ and $\hat{K} = K - dZ$, and $\hat{K}$ 
shares the same good properties with $K$, i.e. satisfies
\bea
\hat{K}_{(0,4)} = &0& = \hat{K}_{(1,3)} \\
d\hat{K}=0.
\eea
It could then be that only for a suitable choice of $Z$ one may 
satisfy \eref{conssst}. The problematic feature of such a solution
lies in the fact that the three superform $Z$ has to be constructed
using explicitly the fields of the tensor multiplet, whose SUSY 
transformations are not known from the beginning. In fact,  the
combinations which involve only the fields appearing in the
parametrizations \eref{T1}-\eref{Tend} have -- for dimensional reasons
--  already been exhausted by our "minimal" decomposition of $trR^2$. 
Therefore, if a solution can be achieved only for a non vanishing
$Z$ it can presumably not be obtained in a closed form.

\subsection{Duality in the presence of a Lorentz-Chern-Simons form}

We suppose in this section that \eref{conssst} holds indeed and analyze 
again the possibility of dualizing the connections $\f$, $A$ and $B$ 
in the non-minimal model constructed in the preceding section.

The answer is very simple and positive for what concerns $\tH$ and 
the (abelian) $\tF$. The B.I.'s \eref{id1} and \eref{id2} are again 
satisfied if one defines $\tF$ exactly as in \eref{conF1}-\eref{conF3} 
while for $\tH$ 
one can again use \eref{conH1}-\eref{conH3} with the only difference that now
\beq
\tH_{abc} = \frac{1}{\phi^2} \frac{1}{3!} \e_{abcdef}\left(\hat{H}^{def} - 
\frac{1}{4\phi} \l_i \G^{def}\l^i + c_1 tr(\c^i \G^{def} \c_i)\right).
\eeq

On the other hand the dilaton has now to be described as a zero-form, 
i.e. as a scalar, since the B.I. for $\tV$ can be no longer 
\eref{id3} in that the r.h.s. of \eref{id3} is not closed anymore.
This is due 
to the modified B.I. for $H$, $d H = c_1 tr F^2 + c_2 tr R^2$, and to the fact
that the
Lorentz curvature $R_a{}^b$ cannot be dualized for its intrinsic 
non-abelian nature.

\section{Some concluding remarks}

The variety of couplings in supergravity theories becomes more and more 
restricted as the dimensionality increases. Eleven-dimensional supergravity 
allows only for the pure SUGRA multiplet and only the minimally
coupled theory has been explicitly constructed \cite{CreJu}.
In ten dimensions there is, in addition to the SUGRA multiplet, the YM 
multiplet
 (in the case of $N = 1$ SUSY) and in this case, in addition to the
minimally coupled theory, there exists also an exact solution
for anomaly free $N = 1$, $D=10$ SUGRA-SYM, i.e. in the presence
of a Lorentz-Chern-Simons form \cite{BPT,catena}.

For these dimensions the supersymmetric dualization of the abelian
connections has been carried out in \cite{CL,LT,Ceder,catena}. The results of 
those papers indicated that the existence of the supersymmetric
duals of the abelian connections constitutes actually a general
feature of all supergravity theories, independently of their
dimensionality, even if a general proof of this statement is still 
missing. The absence of such a general theorem is probably related to the
fact that  Hodge--duality  can not be canonically lifted to 
superspace.

In the present paper we extended this analysis to the six-dimensional
case. The essentially new features of six-dimensional supergravity
theories, w.r.t. 10 and 11 dimensions, are constituted by the appearance of 
matter (hyper- and tensor)-multiplets which enlarge significantly
the possible couplings between pure supergravity and the other
multiplets. Our general analysis revealed, however, that these couplings 
are restricted by certain consistency conditions, which are clearly
satisfied for minimal couplings, and also in the anomaly free non-minimal
theory modulo eq. \eref{conssst} which needs still to be checked.
For what concerns duality we were able to confirm the expectation which emerged
from ten and eleven dimensions: the $B_2$ form and the abelian Maxwell
fields can be dualized in both cases, while the dilaton can be dualized 
only in the minimally coupled theory. The possible relevance of the
dual Maxwell fields has already been noticed \cite{LT,SSn} while possible
applications of the dualized dilaton are still outstanding.

Possible generalizations of these results could be gained in the following
directions. Recently rather general couplings of an arbitrary  number $n_T$ 
of tensor-multiplets have been worked out in \cite{NS}. Since in this case one
 $B_2$ curvature (essentially the one belonging to the SUGRA-multiplet) is 
anti-selfdual, while the remaining $n_T$ ones are self-dual, the $B_2$
 curvatures can no longer be dualized while it should still be possible
to dualize the `dilatons' and the abelian Maxwell fields.

Another interesting point concerns the dualizability of the hyperscalars. 
In a globally supersymmetric (and free) theory they can actually be described
as four-form potentials, as has been shown long time ago in \cite{susy6d}, 
while it seems unlikely that they can be dualized in supergravity due to
the quaternionic structure of the underlying manifold.

\paragraph{Acknowledgments} Work of K.L. was supported by the European 
Commission TMR Programme ERBFMRX-CT96-0045 to which K.L. is associated.

\section*{Appendix: Notations and Conventions}

\renewcommand{\theequation}{A.\arabic{equation}}

We write our symplectic Majorana-Weyl spinors as $\psi_{\a i}$ 
(left-handed) and $\psi^{\a i}$ (right-handed) where $i = 1,2$ is 
an $USp(1)$ index which can be 
raised and lowered with the invariant antisymmetric tensor $\e_{ij}$
\beq
\psi_i = \e_{ij} \psi^j, \qquad \psi^i = \e^{ji} \psi_j,
\eeq

\nin
while $\a$ is a chiral $SO(1,5)$ spinor index which cannot be raised 
or lowered. The symplectic Majorana-Weyl condition reads
\beq
\e^{ij} \psi^{\a j} = O^{\a\b} \psi^{\star \b i} \hbox{ \hfill } 
\label{A1}
\eeq

\nin
where the matrix $O$ satisfies
\beq
O^T = -O, \quad O^\star = O, \quad O^2 = -\hbox{\msbm I}.
\eeq
The matrices $(\G^{a})_{\a\b}$ and $(\G^a)^{\a\b}$ span a Weyl-algebra,
$(\G_{(a})_{\a\b} (\G_{b)})^{\b\g} = \eta_{ab} \delta_\a^\g$,
and satisfy the hermiticity condition
\beq
O \G^a{}^\dagger O = \G^a. \hbox{ \hfill } \label{A2}
\eeq

\nin
Notice, however, that the relations \eref{A1}--\eref{A2} need  
never used explicitly.

The duality relations for the anti-symmetrized $\G$-matrices is
\beq
(\G_{a_1\ldots a_k})_{\a\b} = - (-1)^{k(k+1)/2} \frac{1}{(6-k)!} 
\e_{a_1\ldots a_6} \G^{a_{k+1}\ldots a_6}_{\a\b}
\eeq

\nin
where no "$\g_7$" appears since our $\G$-matrices are $4\times 4$ 
Weyl matrices, and the cyclic identity reads
\beq
(\G^a)_{\a(\b} (\G_a)_{\g)\delta} = 0.
\eeq

Another fundamental identity is
\beq
(\G_a)_{\a\b}(\G^a)^{\g\delta} = -4 \delta^{\g}_{[\a} \delta^\delta_{\b]}.
\eeq

\end{document}